# Angle-Dependent van Hove Singularities and Their Breakdown in Twisted Graphene Bilayers

Wei Yan[1], Lan Meng[1], Mengxi Liu[2], Jia-Bin Qiao[1], Zhao-Dong Chu[1], Rui-Fen Dou[1], Zhongfan Liu[2], Jia-Cai Nie[1], and Lin He[1,*]

[1] Department of Physics, Beijing Normal University, Beijing, 100875, People's Republic of China
[2] Center for Nanochemistry (CNC), College of Chemistry and Molecular Engineering, Peking University, Beijing 100871, People's Republic of China

**The creation of van der Waals heterostructures based on a graphene monolayer and other two-dimensional crystals has attracted great interest because atomic registry of the two-dimensional crystals can modify the electronic spectra and properties of graphene. Twisted graphene bilayer can be viewed as a special van der Waals structure composed of two mutual misoriented graphene layers, where the sublayer graphene not only plays the role of a substrate, but also acts as an equivalent role as the top graphene layer in the structure. Here we report the electronic spectra of slightly twisted graphene bilayers studied by scanning tunneling microscopy and spectroscopy. Our experiment demonstrates that twist-induced van Hove singularities are ubiquitously present for rotation angles $\theta$ less than about 3.5$^o$, corresponding to moiré-pattern periods $D$ longer than 4 nm. However, they totally vanish for $\theta > 5.5^o$ ($D < 2.5$ nm). Such a behavior indicates that the continuum models, which capture moiré-pattern periodicity more accurately at small rotation angles, are no longer applicable at large rotation angles.**

Graphene's novel electronic properties are a consequence of its two-dimensional honeycomb lattice [1]. Its electronic spectra are relatively easy to be tuned because graphene is a single-atom-thick membrane of carbon [2-4]. Very recently, it was demonstrated that a layer of hexagonal boron nitride (hBN) in contact with graphene can generate a periodic potential felt by graphene and lead to profound changes in graphene's electronic spectrum [5-10]. This provides an effective route to control the electronic spectra and properties of graphene via the creation of van der Waals heterostructures [5-10]. Graphene placed on top of another graphene monolayer with stacked misorientation forms a unique two dimensional van der Waals structure, i.e., twisted graphene bilayer [11-21], in which the graphene-on-graphene moiré modifies the electronic spectra [16,17,19]. The period of the moiré pattern $D$ is related to the rotation angle $\theta$ by $D = a/[2\sin(\theta/2)]$ with $a = 0.246$ nm the lattice parameter of graphene. This unique layered structure exhibits many fascinating physical properties beyond that of graphene monolayer due to interlayer coupling [16-20]. For example, the quasiparticles in twisted graphene bilayer are expected to show tunable chirality and adjustable probability of chiral tunneling [20].

At small rotation angles electronic spectra of twisted graphene bilayer have been experimentally demonstrated to follow the predictions of the continuum models [11] and show twist-induced van Hove singularities (VHSs) [13-15,22,23], which directly arise from the finite interlayer coupling. However, the VHSs were not always observed and several experiments indicate that the electronic properties of the twisted graphene bilayer resemble a single graphene sheet [13,14,24-27]. Obviously, the experimental results concerning the electronic spectra of twisted graphene bilayer are lacking of consistency. Theoretically, it is well accepted that the continuum models depict well the moiré periodicity at small rotation angles but ignore some details of the atomic arrangement, which are expected to be important at large rotation angles [12,21,28,29]. To exploit the electronic spectra of twisted graphene bilayer, it is of particular interest to explore down to which length scale of the moiré period the twist-induced VHSs persists, and how it is altered below that. We show that the VHSs are ubiquitously present for $D > 4$ nm ($\theta < 3.5^o$), whereas they totally vanish for $D < 2.5$ nm ($\theta > 5.5^o$) in twisted graphene bilayers grown on Rh foils. Both the cases of presence and absence of the VHSs in the spectra coexist for 2.5 nm $< D <$ 4 nm ($3.5^o < \theta < 5.5^o$).

To explore the angle dependence of electronic spectra, we studied twisted graphene bilayer on Rh foils prepared by chemical vapour deposition [15,19,30]. The thickness of graphene can be tuned by controlling the cooling rate in growth process, and only samples mainly covered with bilayer graphene are further studied in this work. Our experimental result suggests that the vertical directions of the bilayer are randomly rotated and there is no preferable rotation angle between the two layers [30]. The 33 twisted graphene bilayers studied in this work are randomly observed on different regions of several Rh foils. The scanning tunneling microscopy (STM) system was an ultrahigh vacuum four-probe SPM from UNISOKU. All STM and scanning tunneling spectra (STS) measurements were performed at liquid-nitrogen temperature and the images were taken in a constant-current scanning mode. The

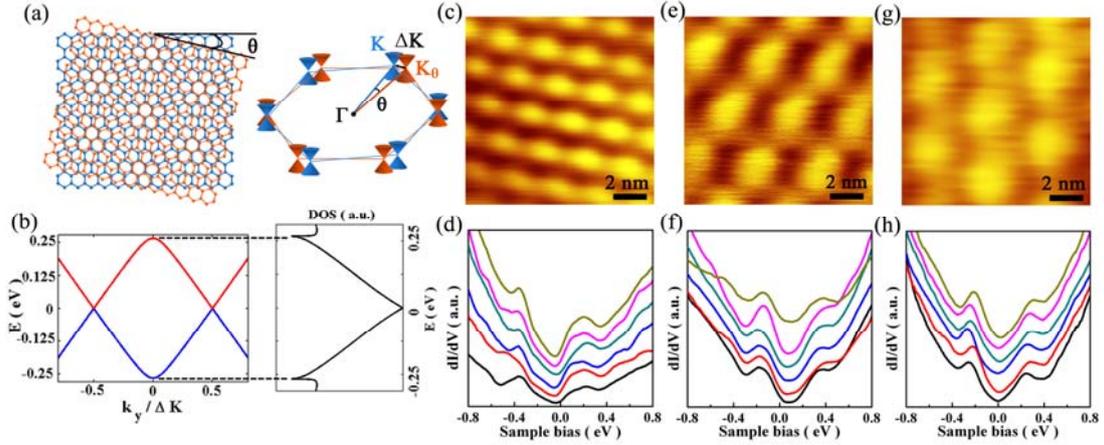

**Figure 1** (color online). (a), Schematic structural model of two misoriented honeycomb lattices with a twist angle $\theta$ (left) and the schematic Dirac cones, K and K$_\theta$, of the two graphene layers in the reciprocal space (right). (b), Electronic band structure of twisted bilayer graphene along a line joining the two Dirac points (left) and the corresponding low-energy density of states calculated numerically according to the well-known formula $\frac{S}{4\pi^2}\oint \frac{1}{|\nabla_k E(\vec{k})|}dk$ (right). In the calculation, $\theta = 3.9°$ and $t_\theta = 0.04$ eV are used. (c), (e), and (g), The 11×11 nm$^2$ STM topographies of graphene bilayers on Rh foils showing three different periods of moiré patterns. (c), $\theta = 5.2°$, $D = 2.7$ nm ($V_{sample} = -0.5$ V, $I = 0.22$ nA). (e), $\theta = 3.9°$, $D = 3.6$ nm ($V_{sample} = 0.7$ V, $I = 0.22$ nA). (g), $\theta = 3.4°$, $D = 4.1$ nm ($V_{sample} = -0.5$ V, $I = 0.19$ nA). (d), (f), and (h) show the tunneling spectra, i.e., d$I$/d$V$-$V$ curves, acquired on different positions of the moiré patterns in panels (c), (e), and (g), respectively. The spectra have been vertically offset for clarity. The two peaks flanking the Dirac point of the spectra are attributed to the twist-induced VHSs of graphene bilayer. The positions of the two VHSs are not always symmetric around the Fermi level, which may arise from different charge transfer between the graphene and the substrate on different samples.

STM tips were obtained by chemical etching from a wire of Pt(80%) Ir(20%) alloys. Lateral dimensions observed in the STM images were calibrated using a standard graphene lattice. The STS spectrum, i.e., the d$I$/d$V$-$V$ curve, was carried out with a standard lock-in technique using a 957 Hz a.c. modulation of the bias voltage (with the modulation amplitude of 20 mV). In order to ascertain the validity of the STS spectra, all the STM tips are calibrated in highly oriented pyrolytic graphite (HOPG) and only the tips acquired the "V-shaped" spectra are used in this work.

It was demonstrated previously that the coupling between graphene and the Rh foil is very weak [31] (also see Fig. S1 of Supplementary Information [32] for the experimental result) and the graphene bilayers on Rh foils have a strong twisting tendency [15,19,30]. Therefore, the graphene bilayer on Rh foils provides an ideal platform for twist engineering of electronic spectra. The twisting not only results in the moiré pattern, but also splits the parabolic spectrum of Bernal graphene bilayer into two Dirac cones in reciprocal space [11,12,20,33], as shown in Fig. 1(a). The resulting relative shift of the Dirac points on the different layers is |Δ**K**| = 2|**K**|sin(θ/2), where **K** is the reciprocal-lattice vector. If there is a finite interlayer coupling between the two graphene sheets, two saddle points appear along the intersection of the two cones and generate two low-energy VHSs in the density of states (DOS) at energies about $E_{\text{VHS}}^{\pm} = \pm(\hbar v_F \Delta K / 2 - t_\theta)$, as shown in Fig. 1(b). Here $v_F$ is the Fermi velocity of the graphene, $t_\theta$ is the interlayer hopping parameter (see Supplementary Information [32] for details of calculation) [13-15,22]. Figure 1(c), (e), and (g) show three typical STM images of the twisted graphene bilayers on Rh foils and their

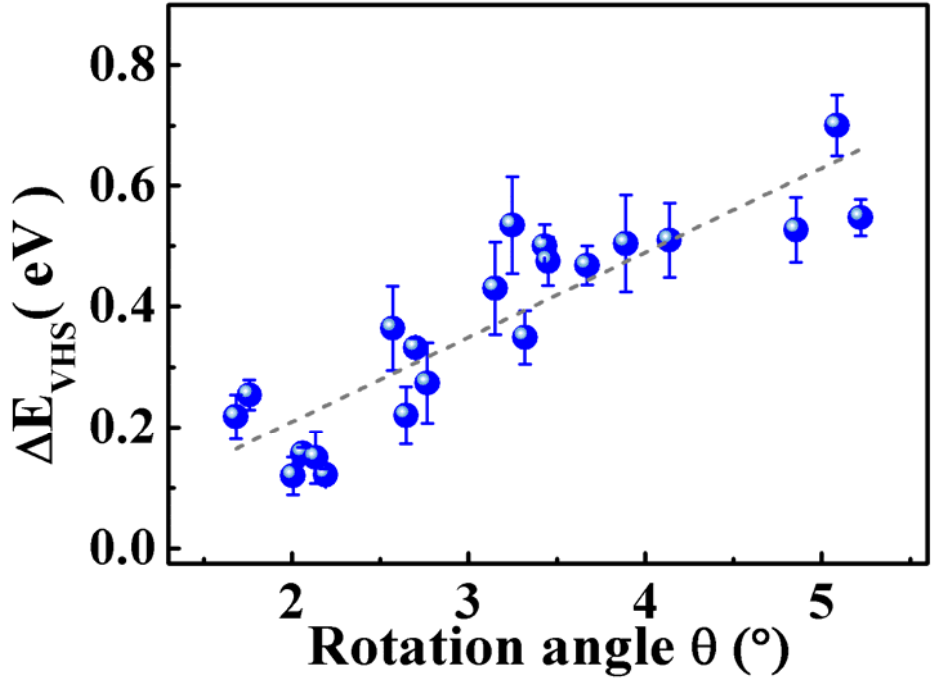

**Figure 2** (color online). Angle-dependent VHSs in the twisted graphene bilayers. Solid blue circles are the average experimental data measured in different twisted graphene bilayers. Error bars in energy represent the standard deviation of the $\Delta E_{VHS}$ observed in our experiment. Gray dot line is a linear fit of the experimental data using the empirical equation $\Delta E_{VHS} = \hbar v_F \Delta K - 2t_\theta$ with $v_F = 0.7 \times 10^6$ m/s and $t_\theta = 0.04$ eV.

corresponding STS. The superstructures in the STM images are attributed to the moiré pattern arising from a stacked misorientation between adjacent graphene layers. The tunneling spectrum gives direct access to the local DOS of the surface. Therefore, two peaks in the STS spectra, as shown in Fig. 1(d), (f), and (h), are attributed to the twist-induced VHSs in graphene bilayers, as reported previously [13-15,22,23]. In order to ascertain the reproducibility of the results, several tens of spectra were measured at different positions of each graphene bilayer (see Fig. S3 in

Supplementary Information [32] for a typical experimental result). All the spectra of the graphene bilayer show similar main features with small variations in the peak spacings and amplitudes of the VHSs, which depend on the positions of the moiré pattern. These variations lead to the observed standard deviation of $\Delta E_{VHS}$ (the energy difference of the two VHSs) in Fig. 2, which will be discussed subsequently.

To further verify the origin of the two peaks in the STS spectra, we carried out measurements on several tens of twisted graphene bilayers with different rotation angles. Figure 2 summarizes the energy difference of the VHSs $\Delta E_{VHS}$ as a function of the rotation angles for 21 graphene bilayers showing the two DOS peaks in the STS spectra (see Fig. S4 for more experimental results and Fig. S5 in Supplementary Information [32] for details in determining the value of $\Delta E_{VHS}$). For small twisted angle, $\sin\theta \sim \theta$, the value of $\Delta E_{VHS}$ is expected to increase with $\theta$ according to $\Delta E_{VHS} \approx \hbar v_F \Delta K - 2t_\theta$ and $|\Delta \mathbf{K}| = 2|\mathbf{K}|\sin(\theta/2)$. The twisting-angle dependence of $\Delta E_{VHS}$ is its unmistakable characteristic [13-15,22,23]. This result further demonstrates that the continuum models [11] capture well the main features of low-energy electronic spectrum in twisted graphene bilayers, at least, for small rotation angles. The wide distribution of the value of $\Delta E_{VHS}$, as shown in Fig. 2, may arise from variations of the interlayer coupling in different twisted bilayers. The observed standard deviation of $\Delta E_{VHS}$ is partially attributed to the variations of the interlayer coupling at different positions of each bilayer. We can obtain the two parameters, i.e., the $v_F$ and $t_\theta$, by fitting the experimental data to $\Delta E_{VHS} \approx \hbar v_F \Delta K - 2t_\theta$. The obtained $v_F$ is slightly lower than $1.0\times10^6$ m/s (the expected value of pristine

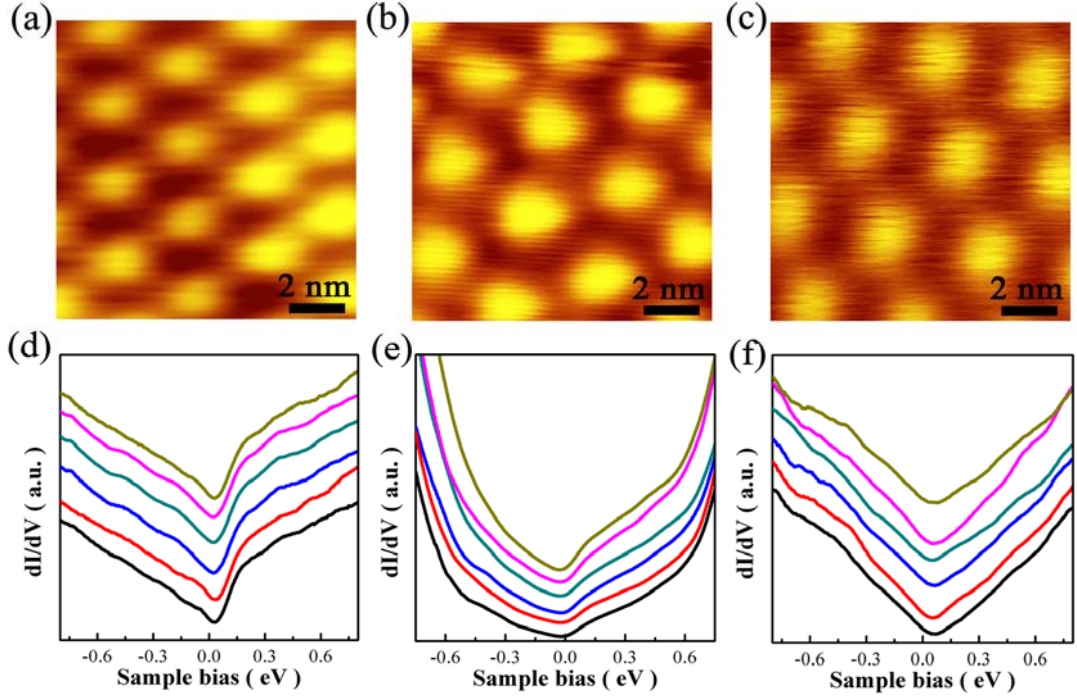

**Figure 3** (color online). (a-c), The 11.6×11.6 nm² STM topographies of three graphene bilayers with different twisted angles. (a), $\theta$ = 4.8°, $D$ = 3.0 nm ($V_{sample}$ = -0.5 V, $I$ = 0.20 nA). (b), $\theta$ = 4.0°, $D$ = 3.5 nm ($V_{sample}$ = 0.37 V, $I$ = 0.24 nA). (c), $\theta$ = 3.7°, $D$ = 3.8 nm ($V_{sample}$ = 0.41 V, $I$ = 0.21 nA). (d-f), Typical d$I$/d$V$-$V$ curves, acquired on the moiré pattern in panels (a-c) respectively. These spectra show no discernible structures and are almost identical to that of graphene monolayer.

graphene monolayer). However, it is very difficult to judge whether the Fermi velocity is reduced or not with decreasing the rotation angle only based on this fitting because of the wide distribution of the data (see Fig. S6 in Supplementary Information [32] for details of discussion about the fitting).

More importantly, not all the twisted graphene bilayers in our experiment exhibit the two VHSs in the tunneling spectra. Figure 3 show three typical examples. The three twisted graphene bilayers exhibit similar moiré patterns as that in Fig. 1,

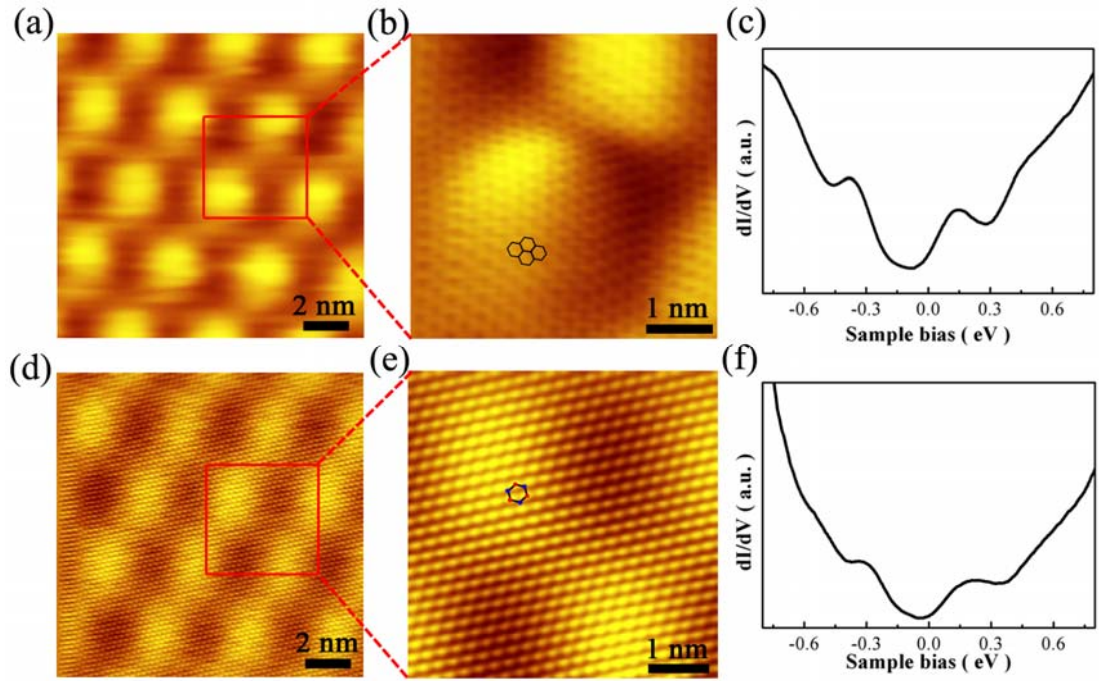

Figure 4 (color online). (a) and (d), 13.4×13.4 nm$^2$ STM images of twisted graphene bilayers grown on Rh foil. (a), $\theta = 3.7°$, $D = 3.84$ nm, $V_{sample} = 0.7$ eV, $I = 0.22$ nA. (d), $\theta = 3.6°$, $D = 3.89$ nm, $V_{sample} = -0.62$ eV, $I = 0.20$ nA. (b) and (e) show zoom-in topographies of the red frames in (a) and (d) with atomic-resolution images of honeycomb lattice and triangular lattice, respectively. The hexagonal structure of graphene is overlaid onto the STM images. (c) and (f), Tunneling spectra, i.e. d$I$/d$V$-$V$ curves, acquired on the moiré pattern in panels (b) and (e) respectively.

whereas their STS spectra are almost identical to that of single-layer graphene [3,4,13,14]. Similar results, which suggest that the electronic properties of twisted graphene bilayers resemble a single graphene sheet, were obtained in twelve bilayers with different rotation angles in our experiments. Different STM tips were used to confirm the reproducibility of the STS spectra and to remove any possibility of tip artefacts (see Fig. S7 of Supplementary Information [32] for more experimental data). Here we should point out that the main features of the STS, as shown in Fig. 3 (and also in Fig. 1), are robust and irrespective of different tips and different positions of each bilayer. Our experimental result indicates that there are two distinct spectra in twisted graphene bilayers. Such a discrepancy, as shown in Fig. 1 and Fig. 3, reveals the main controversy of all current experimental data [13-15,22-26] and understanding [11,12,21,28] on this unique layered system.

Previously, the decoupled behavior of twisted graphene bilayers (i.e., the bilayers did not show the low-energy VHSs in their spectra) was attributed to that the layer pairs may not have been adjacent [12,14]. Naively, we could, therefore, expect to distinguish the electronically decoupled and coupled (i.e., the bilayers show the low-energy VHSs in their spectra) bilayers from their STM images. For *AB* (Bernal) stacked bilayer, atomic resolution usually shows triangular contrast in the STM images because of A/B atoms asymmetry [4,13,27]. A slightly stacked misorientation between the consecutive layers with a finite interlayer coupling is therefore expected to show a seemingly triangular pattern (or very strong A/B atoms asymmetry), and the honeycomb contrast of isolated graphene monolayer should be recovered when the two graphene sheets are electronically decoupled. However, carefully examination of our experimental data indicates that both the honeycomb and triangular patterns can be observed in the coupled twisted bilayers, as shown in Fig. 4, and in the decoupled bilayers, as shown in Fig. S8 of Supplementary Information [32]. We also carefully exclude the effect of tunneling conditions, for example the bias voltage, on the appearance of the graphene bilayers in atomically resolved images, as shown in Fig. S9 [32]. This result implies that we can not distinguish the electronically coupled and

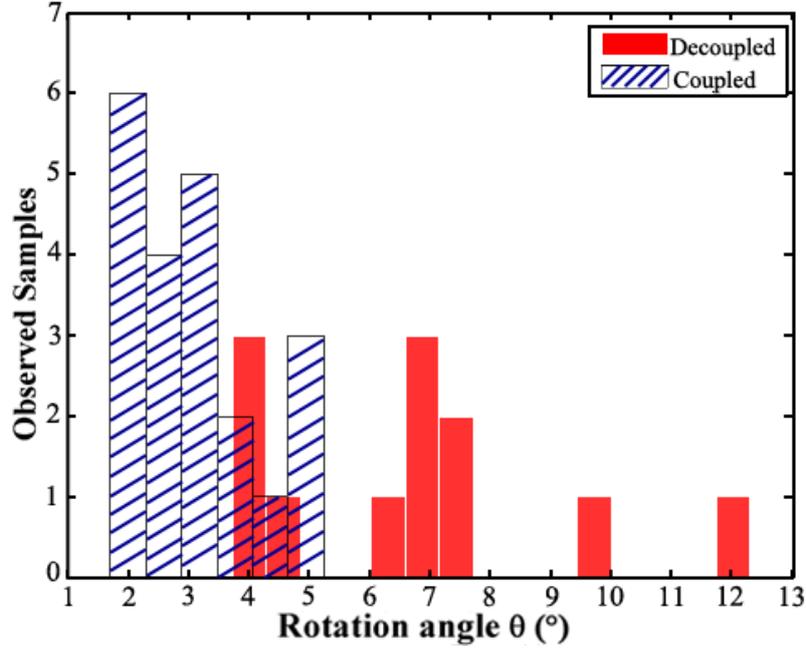

**Figure 5** (color online). The number of coupled graphene bilayer (opened bars with blue lines) and decoupled graphene bilayer (red bars) versus the rotation angles from our experiments.

decoupled bilayers simply only based on their STM images. Obviously, the previous interpretation can not fully explain the results reported here.

To further explore the origin of the two distinct types of spectra observed in the twisted graphene bilayers, we plot the number of the coupled bilayers and the decoupled bilayers versus the rotation angles from our experiments in Fig. 5. According to the result shown in Fig. 5, it appears immediately that whether the two layers are electronically coupled or decoupled depends on the rotation angles sensitively. For $\theta < 3.5°$ ($D > 4$ nm), the VHSs are ubiquitously observed, whereas the electronic spectra of the two graphene sheets always resemble a single graphene sheet for $\theta > 5.5^o$ ($D < 2.5$ nm). In the intermediate case, i.e., $3.5^o < \theta < 5.5^o$ (2.5 nm $< D <$ 4 nm), both the coupled and decoupled cases coexist. This result indicates that the continuum models, which capture moiré-pattern periodicity more accurately at small rotation angles, are no longer applicable at large rotation angles. Such a result is reasonable since that the atomic registry of small moiré pattern is expected to play an important role in determining the electronic band structures of twisted graphene bilayers [21]. The decoupled between adjacent graphene sheets with a separation of about 0.35 nm was identified as a destructive interference between the layers [21]. The existence of the intermediate region between the coupled region and decoupled region, as shown in Fig. 5, may arise from the variations of the interlayer coupling in different twisted bilayers. Here we should point out that the boundary between the coupled region and decoupled region may depend on the substrate of twisted graphene bilayers (the substrate may either strengthen or weaken the interlayer coupling). For twisted graphene bilayers with large interlayer coupling, the boundary between the two regions is expected to appear at relative larger twisted angles. Further experiments of twisted graphene bilayers on other substrates should be carried out to clarify the origin of the intermediate region and the effect of substrate. Additionally, more theoretical studies are cried for to quantitative understand the boundary of the coupled region and decoupled region.

In summary, our results presented here demonstrate that twisted graphene bialyer is a unique layered structure with tunable electronic spectra. The electronic properties of this layered system can be easily tailored by misorientation of the layers even with an atomic scale of layer separation. This observation may pave the way for twisting engineering of electronic properties in the two-dimensional van der Waals structures.

**Acknowledgements**

This work was supported by the Ministry of Science and Technology of China (Grants Nos. 2014CB920903, 2013CBA01603, 2013CB921701), the National Natural Science Foundation of China (Grant Nos. 11374035, 11004010, 51172029, 91121012), the program for New Century Excellent Talents in University of the Ministry of Education of China (Grant No. NCET-13-0054), and Beijing Higher Education Young Elite Teacher Project (Grant No. YETP0238). W. Yan and L. Meng contributed equally to this paper.

*Email: helin@bnu.edu.cn